\documentclass[10pt,aps,prb,twocolumn,floatfix,showpacs,superscriptaddress,longbibliography]{revtex4-1}

\usepackage{graphicx,tabularx}
\usepackage{amsmath,amsfonts,dsfont}
\usepackage{bm}
\usepackage[colorlinks=true, citecolor=blue, linkcolor=red, urlcolor=blue]{hyperref}
\usepackage[all]{hypcap}

\graphicspath{{figures/}}

\newcommand{\BiSe}{\mathrm{Bi}_2\mathrm{Se}_3}
\newcommand{\sign}{\mathrm{sign}}
\newcommand{\re}{\mathrm{Re}\,}

\newcommand{\angstrom}{\textup{\AA}}

\newcolumntype{C}{>{\centering\arraybackslash}X}

\begin{document}

\title{Crystalline topological states at a topological insulator junction}
\author{C. De Beule}
\email{christophe.debeule@uantwerpen.be}
\affiliation{Department of Physics, University of Antwerp, 2020 Antwerp, Belgium}
\author{R. Saniz}
\affiliation{Department of Physics, University of Antwerp, 2020 Antwerp, Belgium}
\author{B. Partoens}
\affiliation{Department of Physics, University of Antwerp, 2020 Antwerp, Belgium}
\begin{abstract}
We consider an interface between two strong time-reversal invariant topological insulators having surface states with opposite spin chirality, or equivalently, opposite mirror Chern number. We show that such an interface supports gapless modes that are protected by mirror symmetry. The interface states are investigated with a continuum model for the $\BiSe$ class of topological insulators that takes into account terms up to third order in the crystal momentum, which ensures that the model has the correct symmetry. The model parameters are obtained from \emph{ab initio} calculations. Finally, we consider the effect of rotational mismatch at the interface, which breaks the mirror symmetry and opens a gap in the interface spectrum.
\end{abstract}

\maketitle

\section{Introduction}

Time-reversal invariant topological insulators (TIs) are bulk insulators that have metallic surface states on any surface. Moreover, the surface states are topologically protected by time-reversal (TR) symmetry and charge conservation \cite{Kane2005a,Bernevig2006a,Konig2007,Fu2007a,Fu2007b,Hsieh2008}. Topological insulators are characterized by a $\mathbb Z_2$ topological invariant that corresponds to the parity of the number of Dirac points enclosed by the surface Fermi surface through the bulk-boundary correspondence \cite{Hasan2010,Hasan2011,Qi2011,Ando2013a}. In the topologically nontrivial case, the number of surface Dirac points is odd since perturbations that respect TR symmetry can only pairwise annihilate Dirac points. Hence, the topological surface state is given by a single Dirac cone in the simplest case. The surface states are protected against weak localization since TR symmetry forbids elastic backscattering of the topological surface states from nonmagnetic scatterers. Furthermore, there also exist weak topological materials, in the sense that they are not robust against disorder. Topological crystalline insulators have surface states that are protected by crystalline symmetries \cite{Fu2011,Ando2015}. For example, SnTe has gapless surface states protected by mirror symmetry \cite{Hsieh2012,Tanaka2012}. The topological phase of SnTe is characterized by the mirror Chern number $n_{\mathcal M}=-2$, leading to a pair of surface Dirac cones on any surface that preserves the mirror symmetry.

Topological crystalline states can also occur in heterostructures of TIs with mirror symmetry. For these TIs, the surface state survives even if TR is broken as long as mirror symmetry is preserved. This occurs, for example, if there is a magnetic field along the mirror axis. For the $\BiSe$ class of TIs, $n_{\mathcal M}=\pm1$, where the sign determines the spin chirality of the surface states (handedness of the spin texture) and the absolute value gives the number of surface Dirac cones \cite{Rauch2014,Zhang2009,Hsieh2009a,Teo2008}. Hence, gapless modes should also exist at any mirror-symmetric interface between TIs with opposite spin chirality since this corresponds to a change $\Delta n_{\mathcal M} = 2$ \cite{Takahashi2011,Apalkov2012,DeBeule2013,Habe2013a}.

There are already examples in the literature on how the sign of the spin chirality can be tuned. The spin chirality is determined by the sign of the spin-orbit coupling (SOC) constant \cite{Liu2010,Rauch2015}. For isolated atoms, the SOC constant is always positive because the potential is always attractive. However, in cubic binary materials, such as HgS and strained HgTe, which are also TIs, the spin-orbit splitting can be effectively negative due to contributions from $d$ orbitals \cite{Vidal2012}. In strained HgTe, this contribution is too small and $n_{\mathcal M }= -1$ \cite{Brune2011,Rauch2015}. In HgS, however, the $p$-$d$ hybridization leads to an effective negative SOC constant for $p$ orbitals and therefore $n_{\mathcal M }= +1$ \cite{Virot2011,Rauch2015}. More generally, strained HgTe$_x$S$_{1-x}$ has been shown to exhibit topological phase transitions between strong TIs with $n_{\mathcal M} = \pm1$ as a function of the strain and the composition $x$, with the limiting cases of strained HgTe $(n_{\mathcal M } = -1)$ and unstrained HgS $(n_{\mathcal M }= +1)$ \cite{Rauch2015}. Heterostructures of HgTe$_x$S$_{1-x}$ where the strain and $x$ are tuned accordingly could therefore be a possible experimental realization. 
It is clear that the spin chirality depends strongly on the orbital character of the surface state.
In the $\BiSe$ family of TIs, orbitals normal to the surface favor a clockwise $(n_{\mathcal M} = -1)$ spin chirality while in-plane orbitals favor an anticlockwise spin chirality $(n_{\mathcal M} = +1)$ \cite{Cao2013,Zhang2013}. Hence, since the inverted bands at the $\Gamma$ point are mostly $p_z$, the spin texture of the surface states depends strongly on the orientation of the surface. It should therefore in principle be possible to engineer the spin chirality of the topological surface state in $\BiSe$-like TIs.

In this paper, we use an effective continuum model to investigate the interface between TIs of the $\BiSe$ class with opposite spin chirality. The model parameters are obtained from fitting the energy bands to \emph{ab initio} calculations. The $\BiSe$ class of TIs have layered crystals structures that consist of stacked quintuple layers, given by Se1-Bi-Se2-Bi'-Se1' for $\BiSe$, where each quintuple layers consists of five atomic layers with trigonal symmetry that are ABC stacked \cite{referee2}. Moreover, an interface between these materials parallel to the layers preserves the mirror symmetry. 

The paper is further organized as follows: in Sec.\ \ref{sec:model} we introduce the model where we discuss the symmetries and derive the general solution for states confined in the direction perpendicular to the interface. Then, we implement the boundary conditions by imposing continuity of the probability current density at the interface. We show our results for the gapless interface states in our proposed setup in Sec.\ \ref{sec:results} and discuss how they are protected by mirror symmetry, both in terms of scattering of the surface state at the interface and bulk topology by calculating the mirror Chern number. Furthermore, we calculate the mirror eigenvalues to explicitly show that the gapless states are protected by mirror symmetry. Finally, we discuss the effect of rotational mismatch between the two materials which breaks the mirror symmetry and opens a gap. We present the summary and conclusions of the paper in Sec.\ \ref{sec:summary}.

\section{Model}
\label{sec:model}

First, we discuss the low-energy model of $\BiSe$, which also applies to other topological insulators (TIs) with the same crystal structure. For the $\BiSe$ class of TIs, there is a single band inversion at the origin $\Gamma$ of the Brillouin zone \cite{Zhang2009,Xia2009}. Hence, it is sufficient to consider only bands near $\Gamma$ to understand the topological properties. At the $\Gamma$ point, the bands near the Fermi level are spanned by four states with angular momentum $m_j = \pm 1/2$ and parity $\mathcal P = \pm$ \cite{Liu2010}. States with $\mathcal P = \pm$ arise from hybridization between the 6p (Bi) and 4p (Se) valence orbitals. Because of the large energy difference between these orbitals, the hybridized states are mostly localized on Bi ($\mathcal P = +$) and (outer) Se ($\mathcal P = -$) atoms \cite{Liu2010}. On the other hand, states with $m_j = \pm 1/2$ are spin-orbit coupled superpositions of $\left| p_z \uparrow \right>$ with $\left| p_+ \downarrow \right>$ and $\left| p_z \downarrow \right>$ with $\left| p_- \uparrow \right>$, respectively. However, since the crystal-field splitting is much stronger than the spin-orbit coupling, these states are mainly $p_z$, so that $m_j$ is proportional to the electron spin. Therefore, the Hilbert space of the model is approximately spanned by $p_z$ orbitals $\{ \left| \mathrm{Bi} \uparrow \right>, \left| \mathrm{Se} \uparrow \right>, \left| \mathrm{Bi} \downarrow \right>, \left| \mathrm{Se} \downarrow \right> \}$. 

Taking into account the symmetries of $\BiSe$, the effective Hamiltonian becomes \cite{Zhang2009,Liu2010}
\begin{equation}
\label{eq:ham}
H = H_0 + H_3, 
\end{equation}
where
\begin{align}
\begin{split}
H_ 0 & = \varepsilon(\bm k, k_z) \\ & + \mathcal M(\bm k, k_z) \tau_z + \left( A_1 k_z \sigma_z + A_2 \bm k \cdot \bm \sigma \right) \tau_x,
\end{split} \label{eq:ham0} \\
H_3 & = \frac{R_1}{2} \left( k_+^3 + k_-^3 \right) \tau_y + \frac{R_2}{2i} \left( k_+^3 - k_-^3 \right) \sigma_z \tau_x.
\end{align}
Here $\bm k = k_x \bm e_x + k_y \bm e_y$, $k_\pm = k_x \pm i k_y$, and the Pauli matrices $\bm \sigma$ and $\bm \tau$ act on the $m_j = \pm 1/2$ and the parity $\mathcal P = \pm$ (Bi and Se) subspaces, respectively. We also defined
\begin{align}
\varepsilon(\bm k, k_z) & = \varepsilon_0 + \varepsilon_1 k_z^2 + \varepsilon_2 k^2, \label{eq:ch1eps} \\
\mathcal M(\bm k, k_z) & = M - B_1 k_z^2 - B_2 k^2,
\end{align}
where $k = | \bm k |$. Here, we have used the following coordinate system: the $xy$ plane is parallel and the $z$ direction is normal to the quintuple layers, respectively. Moreover, we have taken the $yz$ plane as one of three equivalent mirror planes of $\BiSe$, which are related to each other by threefold rotations around the $z$ axis, so that one of the three mirror axes lies along the $x$ direction. The model parameters $\varepsilon_0$, $\varepsilon_1$, $\varepsilon_2$, $A_1$, $A_2$, $B_1$, $B_2$, $R_1$, $R_2$, and $M$ are obtained from fitting the energy bands \eqref{eq:bulkbands} to \emph{ab initio} calculations and are given in Table~\ref{tab:parameters}. The bulk energy spectrum is given in Appendix \ref{appendix:bands}.

The minimal model for a $\BiSe$-like topological insulator is given by $H_0$, which contains all possible terms up to quadratic order in $\bm k$ and $k_z$. Note that $H_0$ describes an insulator only when $B_{1,2}^2 > \varepsilon_{1,2}^2$. The model describes a strong topological insulator with a single band inversion at the $\Gamma$ point when $M/B_{1,2} > 0$. In this case the character of the valence and conduction band is interchanged between zero and large momentum \cite{Zhang2009}. However, $H_0$ preserves the full rotation symmetry around the $z$ direction. Hence, we include $H_3$ which is given by all cubic terms that reduce the full rotation symmetry to the threefold rotation symmetry ($C_3$) around the $z$ axis of $\BiSe$ \cite{Referee4a,Referee4b}. This excludes terms of order $k_z^3$ since they do not break the full rotation symmetry around $z$.

The total Hamiltonian \eqref{eq:ham} has the following symmetries: time-reversal symmetry which is expressed as $T H(-\bm k, -k_z) T^{-1} = H(\bm k, k_z)$ with $T = i\sigma_y K$ where $K$ is complex conjugation and inversion symmetry $\tau_z H(-\bm k, -k_z) \tau_z = H(\bm k, k_z)$. Combined, time-reversal symmetry and inversion symmetry enforce doubly-degenerate energy bands. We also have threefold rotation around the $z$ axis which is expressed as $e^{-i \frac{\theta}{2} \sigma_z} H(R(-\theta) \bm k,k_z) e^{i \frac{\theta}{2} \sigma_z} = H(\bm k,k_z)$ with $\theta = 2\pi/3$ where $R(\theta)$ is the rotation matrix for a clockwise rotation in the $xy$ plane. Finally, there are three mirror planes related by $C_3$ for which we take the $yz$ plane as the representative giving $\mathcal M_x H(-k_x, k_y, k_z) \mathcal M_x^\dag = H(k_x, k_y, k_z)$ where $\mathcal M_x = -i s_x = -i \sigma_x \tau_z$ with $s_x = \sigma_x \tau_z$ the (dimensionaless) spin operator in the $x$ direction \cite{Silvestrov2012}. 
\begin{table}
	\centering
	\begin{tabularx}{.8\linewidth}{X C | X C}
	\hline \hline
	\multicolumn{2}{c|}{Bulk parameters} & \multicolumn{2}{c}{Surface parameters} \\
	\hline
	$M~(\mathrm{eV})$ & $0.23$ & $\tilde A_2~(\mathrm{eV} \angstrom)$ & 2.29 \\
	$\varepsilon_1~(\mathrm{eV} \angstrom^2)$ & $-0.22$ & $\tilde D~(\mathrm{eV} \angstrom^2)$ & 23.2 \\
	$\varepsilon_2~(\mathrm{eV} \angstrom^2)$ & $20.5$ & $\tilde R_1~(\mathrm{eV} \angstrom^3)$ & 220 \\
	$A_1~(\mathrm{eV} \angstrom)$ & $1.66$ & & \\
	$A_2~(\mathrm{eV} \angstrom)$ & $2.30$ & & \\
	$B_1~(\mathrm{eV} \angstrom^2)$ & $2.19$ & & \\
	$B_2~(\mathrm{eV} \angstrom^2)$ & $27.0$ & & \\
	$R_1~(\mathrm{eV} \angstrom^3)$ & $221$ & & \\
	$R_2~(\mathrm{eV} \angstrom^3)$ & $-294$ & & \\
	\hline \hline
	\end{tabularx}
	\caption{Parameter values for $\BiSe$ from fitting the surface Dirac cone and the bulk bands to \textit{ab initio} calculations (details of the fitting procedure are given in the text).}
	\label{tab:parameters}
\end{table}

\subsection{Probability current density}

The boundary conditions at the interface are found by requiring that the probability current density normal to the interface is continuous. The probability current density $\bm j$ is defined through the continuity equation
\begin{equation}
\partial_t \rho + \nabla \cdot \bm j = 0,
\end{equation}
where $\rho(\bm r, z, t) = \Psi^\dag \Psi$ is the probability density. The first term can be written as
\begin{align}
i \partial_t \rho = i \partial_t \left( \Psi^\dag \Psi \right) & = \Psi^\dag ( i \partial_t \Psi ) - ( i \partial_t \Psi )^\dag \Psi \\
& = \Psi^\dag \hat H \Psi - c.c., \label{eq:derp}
\end{align}
where $\hat H = H(\hat{\bm k}, \hat k_z)$ and we made use of the time-dependent Schr\"odinger equation, $i \partial_t \Psi = \hat H \Psi$ \cite{Messiah1961}. Since we are only concerned with $j_z$, we only consider the terms $\sigma_z \tau_x \hat k_z$, $\tau_z \hat k_z^2$, and $\hat k_z^2$ from $H_0$ given in \eqref{eq:ham0}. We find
\begin{equation}
\Psi^\dag \sigma_z \tau_x \hat k_z \Psi =  \hat k_z ( \Psi^\dag \sigma_z \tau_x \Psi ) + ( \sigma_z \tau_x \hat k_z \Psi )^\dag \Psi,
\end{equation}
where we made use of $\hat k_z \Psi^\dag = - ( \hat k_z \Psi )^\dag$. Hence,
\begin{equation}
\Psi^\dag \sigma_z \tau_x \hat k_z \Psi - c.c. =  \hat k_z ( \Psi^\dag \sigma_z \tau_x \Psi ).
\end{equation}
Likewise, we obtain
\begin{align}
\begin{split}
& \Psi^\dag \tau_z \hat k_z^2 \Psi \\
& \quad = \hat k_z ( \Psi^\dag \tau_z \hat k_z \Psi ) + ( \tau_z \hat k_z \Psi )^\dag \hat k_z \Psi
\end{split} \\
& \quad = \hat k_z [ \Psi^\dag \tau_z \hat k_z \Psi + ( \tau_z \hat k_z \Psi )^\dag \Psi ] + ( \tau_z \hat k_z^2 \Psi )^\dag \Psi,
\end{align}
and therefore
\begin{equation}
\Psi^\dag \tau_z \hat k_z^2 \Psi - c.c. = \hat k_z \left[ 2\,\mathrm{Re} ( \Psi^\dag \tau_z \hat k_z \Psi ) \right].
\end{equation}
Equation \eqref{eq:derp} can then be written as
\begin{equation}
i\partial_t \rho = \hat k_x j_x + \hat k_y j_y + \hat k_z j_z,
\end{equation}
where
\begin{equation}
\label{eq:current}
j_z = \mathrm{Re} \left\{ \Psi^\dag \left[ 2 \left( \varepsilon_1 - B_1 \tau_z \right) \hat k_z + A_1 \sigma_z \tau_x \right] \Psi \right \}.
\end{equation}
Note that $j_z$ contains up to first-order derivatives in $z$ because we do not consider terms of order $k_z^3$ in \eqref{eq:ham}.
\begin{figure}
\centering
\includegraphics[width=\linewidth]{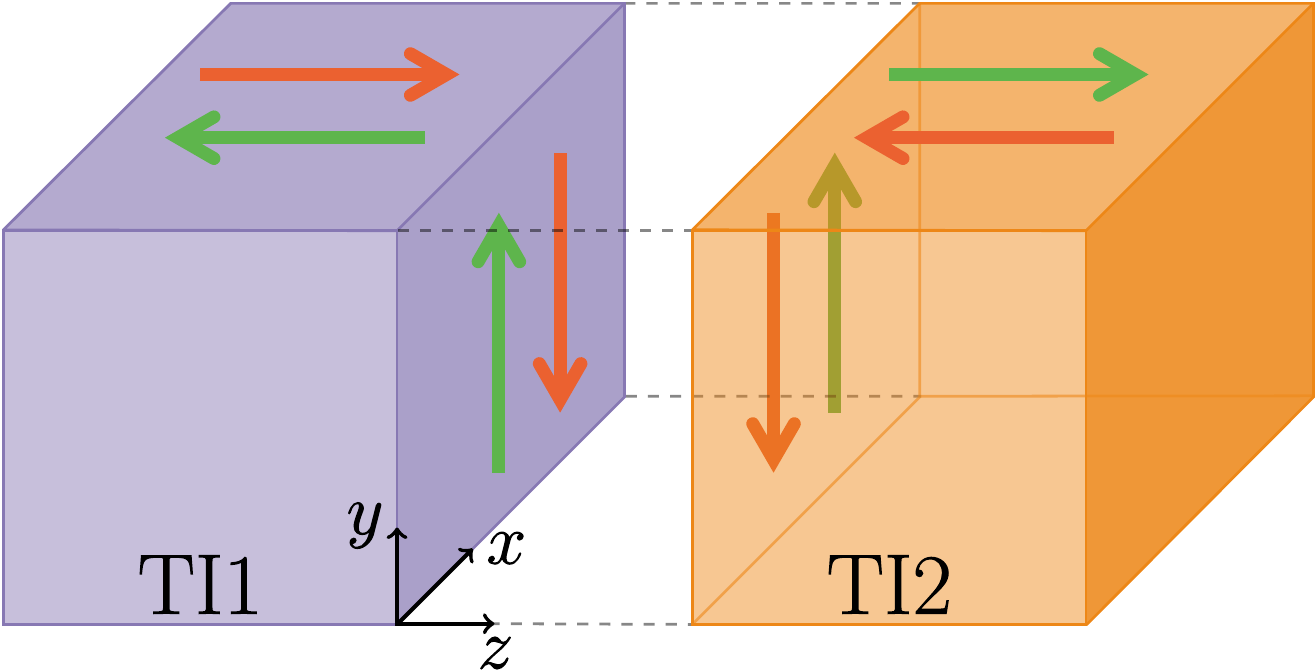}
\caption{Interface between TI1 and TI2 whose topological surface states have opposite spin chirality. The surface states are represented on the $xz$ and $xy$ surface for $k_x=0$ by arrows where the color corresponds to $s_x = \pm$ (green and red). In the figure, TI1 and TI2 have been separated to more clearly show the surface states on the $xy$ surfaces of TI1 and TI2.}
\label{fig:junction}
\end{figure}

\subsection{General solution for localized modes}
\label{sec:localized}

Since we are looking for solutions that are confined in the $z$ direction, we try the \textit{ansatz} 
\begin{equation}
\psi (\bm r, z) = \phi_\lambda e^{\lambda z} e^{i \bm k \cdot\bm r},
\end{equation}
where $\bm r = x \bm e_x + y \bm e_y$. Inserting this trial solution in the Schr\"odinger equation, $H(\hat{\bm k}, \hat k_z) \psi = E \psi$, where $\hat{\bm k} = -i \nabla_{\bm r}$ and $\hat k_z = -i \partial_z $, we obtain
\begin{equation}
\label{eq:secular}
\left[ H ( \bm k , -i \lambda ) - E \right] \phi_\lambda = 0,
\end{equation}
which has a nontrivial solution for $\left| H ( \bm k , -i \lambda ) - E \right| = 0$. This yields an equation for the roots of the square of a depressed quartic equation in $\lambda$ which is given by
\begin{equation}
\label{eq:lambda}
\begin{split}
& D_1 D_2 \lambda^4 + \left[ A_1^2 + D_1 \left( E - L_2 \right) + D_2 \left( E - L_1 \right) \right] \lambda^2 \\
& + i A_1 \left( N_1 + N_2 \right) \lambda + \left( E - L_1 \right) \left( E - L_2 \right) \\
& - A_2^2 k^2 - N_1 N_2 = 0,
\end{split}
\end{equation}
where
\begin{align}
D_{1,2} & = \varepsilon_1 \mp B_1, \\
L_{1,2}(k) & = \varepsilon_0  \pm M + \left( \varepsilon_2 \mp B_2 \right) k^2, \\
N_{1,2}(\bm k) & = k^3 \left( R_2 \sin 3 \theta_{\bm k} \pm i R_1 \cos 3 \theta_{\bm k} \right)
\end{align}
with $\theta_{\bm k} = \arctan \left( k_y / k_y \right)$. Equation \eqref{eq:lambda} gives four distinct $\lambda$ in general, denoted as $\lambda_\alpha(\bm k, E)$ ($\alpha = 1,2,3,4$) which are doubly degenerate. Moreover, if $\lambda_\alpha$ is a solution of \eqref{eq:lambda} then $-\lambda_\alpha^*$ is also a solution. Hence, if there are no imaginary solutions (in which case there would be no normalizable solutions), we can label the $\lambda_\alpha$ such that $\re \lambda_{1,2} > 0$ and $\re \lambda_{3,4} < 0$. The explicit expressions for the $\lambda_\alpha$ are given in Appendix \ref{appendix:lambda}. The corresponding eigenvectors are found from \eqref{eq:secular} and can be written as
\begin{align}
\phi_{\alpha1} & = \begin{bmatrix} -i A_1 \lambda_\alpha + N_2 \\ E -  L_1 + D_1 \lambda_\alpha^2 \\ A_2 k_+ \\ 0 \end{bmatrix}, \label{eq:eigs1} \\
\phi_{\alpha2} & = \begin{bmatrix} 0 \\ A_2 k_- \\ E - L_2 + D_2 \lambda_\alpha^2 \\ +i A_1 \lambda_\alpha - N_2 \end{bmatrix}. \label{eq:eigs2}
\end{align}
The general solution is given by $\Psi(\bm r, z) = \Phi(z) e^{i \bm k \cdot \bm r}$ with
\begin{equation}
\label{eq:general}
\Phi(z) = \sum_{\alpha=1}^4 \sum_{\beta=1}^2 C_{\alpha\beta} \phi_{\alpha\beta} e^{\lambda_\alpha z},
\end{equation}
where the coefficients $C_{\alpha\beta}(\bm k,E)$ are determined by the boundary conditions and the normalization.

\subsection{Topological insulator junction}

We consider an interface ($z=0$) in the $xy$ plane between two TIs which we denote as TI1 $(z<0)$ and TI2 $(z>0)$ which is illustrated in Fig.~\ref{fig:junction}. The first two boundary conditions are given by the normalization condition:
\begin{equation}
\label{eq:boundary1}
\Phi^{(1)}(z\rightarrow-\infty)=0, \qquad \Phi^{(2)}(z\rightarrow+\infty)=0,
\end{equation}
where $\Phi^{(1)}$ and $\Phi^{(2)}$ are the general solutions given in \eqref{eq:general} that correspond to TI1 and TI2, respectively. It follows that the coefficients $C_{\alpha\beta}^{(1)}$ with $\re \lambda_\alpha^{(1)} < 0$ and $C_{\alpha\beta}^{(2)}$ with $\re \lambda_\alpha^{(2)} > 0$ vanish. The other two boundary conditions guarantee the continuity of the $z$ component of the probability current density:
\begin{align}
\Phi^{(1)}(z=0) & = \Phi^{(2)}(z=0), \label{eq:ch1bc1} \\
\left. j_z^{(1)}(\hat k_z) \Phi^{(1)}(z) \right|_{z=0} & = \left. j_z^{(2)}(\hat k_z) \Phi^{(2)}(z) \right|_{z=0}, \label{eq:ch1bc2}
\end{align}
with
\begin{equation}
j_z^{(n)}(\hat k_z) = 2 \big( \varepsilon_1^{(n)} - B_1^{(n)} \tau_z \big) \hat k_z + A_1^{(n)} \sigma_z \tau_x,
\end{equation}
where $n=1,2$ for TI1 and TI2, respectively. In case the parameters $\varepsilon_1$, $B_1$, and $A_1$ of the two TIs are equal, condition \eqref{eq:ch1bc2} reduces to the continuity of the derivative of the wave function. The general solutions become
\begin{alignat}{2}
\Phi^{(1)}(z) & = \sum_{\alpha=1}^2 \sum_{\beta=1}^2 C_{\alpha\beta}^{(1)} \phi_{\alpha\beta}^{(1)} e^{\lambda_\alpha^{(1)} z} && \qquad z < 0, \\
\Phi^{(2)}(z) & = \sum_{\alpha=3}^4 \sum_{\beta=1}^2 C_{\alpha\beta}^{(2)} \phi_{\alpha\beta}^{(2)} e^{\lambda_\alpha^{(2)} z} && \qquad z > 0,
\end{alignat}
where the $\phi_{\alpha\beta}^{(n)}$ are given by \eqref{eq:eigs1} and \eqref{eq:eigs2} and the $\lambda_\alpha^{(n)}$ are obtained from \eqref{eq:lambda} with the corresponding parameters for both TIs. The boundary conditions \eqref{eq:ch1bc1} and \eqref{eq:ch1bc2} become
\begin{widetext}
\begin{equation}
\label{eq:interface}
\begin{vmatrix}
\phi_{11}^{(1)} & \phi_{12}^{(1)} & \phi_{21}^{(1)} & \phi_{22}^{(1)} & -\phi_{31}^{(2)} & -\phi_{32}^{(2)} & -\phi_{41}^{(2)} & -\phi_{42}^{(2)} \\
j_1^{(1)} \phi_{11}^{(1)} & j_1^{(1)} \phi_{12}^{(1)} & j_2^{(1)} \phi_{21}^{(1)} & j_2^{(1)} \phi_{22}^{(1)} & -j_3^{(2)} \phi_{31}^{(2)} & -j_3^{(2)} \phi_{32}^{(2)} & -j_4^{(2)} \phi_{41}^{(2)} & -j_4^{(2)} \phi_{42}^{(2)}
\end{vmatrix} = 0,
\end{equation}
\end{widetext}
where $j_\alpha^{(n)} = j_z^{(n)} \big[ -i\lambda_\alpha^{(n)} \big]$. This equation has no analytical solution and must be solved numerically on a $(\bm k,E)$ grid. Taking time reversal, $C_3$, and mirror symmetry into account we can limit our grid to one $\pi/6$ slice of the Brillouin zone, for example by taking $\theta_{\bm k} \in \left[ \pi/2, 2\pi/3 \right]$.

\subsection{Computational methods}

Here, we discuss the fitting procedure that we used to obtain the model parameters of \eqref{eq:ham} from \emph{ab initio} calculations. Although parameters for $\BiSe$ are available in the literature \cite{Zhang2009,Liu2010}, they do not reproduce the topological surface state very well. Since the interface states arise from the hybridization of the surface states of TI1 and TI2, their properties depend crucially on the correct form for the topological surface state.

The energy bands for the fitting procedure were obtained via {\it ab initio} calculations performed using the VASP package \cite{Kresse1993,Kresse1996}. Electron-ion interactions were described using the projector augmented-wave (PAW) potentials \cite{Bloechl1994,Kresse1999}. 
As previous authors, we treated the Bi $6s6p$ and Se $4s4p$ as valence electrons \cite{Govaerts2014,Zhao2017}, and the experimental structural parameters were used for the calculations (see Table~\ref{tab-a})  \cite{Madelung1998}. The plane wave basis set cutoff was set to 300 eV and spin-orbit coupling was included. Total energies were converged to within $10^{-5}$~eV, using the Perdew-Burke-Ernzerhof (PBE) exchange and correlation functional \cite{Perdew1996}. We first obtained a well converged density using a $6\times 6\times 6$ grid for the Brillouin zone integrations. Then we calculated the eigenvalues on a fine mesh of $21\times 21\times 21$ points on a cubic region of side $0.1~\angstrom^{-1}$ around the $\Gamma$ point. The surface states were calculated using a four quintuple-layer slab with a $21~\angstrom$ vacuum layer to prevent slab-slab interactions. The computational parameters are the same as for the bulk calculation, except that the 2D Brillouin zone integrations were done using a $8\times8$ grid. The surface Dirac cone was then obtained with a fine mesh of $21\times21$ points on a square region of side $0.1~\angstrom^{-1}$ around the $\bar \Gamma$ point.

The model parameters were obtained by fitting the surface Dirac cone and the bulk energy bands from the \emph{ab initio} calculations with a least-squares method. First, we fitted the surface Dirac cone from the slab calculation to
\begin{equation}
E_s^\pm(\bm k) = \tilde \varepsilon_0 + \tilde D k^2 \pm \sqrt{( \tilde A_2 k )^2 + ( \tilde R_1 k^3 \cos 3\theta_{\bm k} )^2},
\end{equation}
which is obtained from \eqref{eq:ham} by perturbation theory on the exact solution for the surface state at $k=0$ \cite{Liu2010,Shan2010}. The parameters for the surface state are related to the bulk parameters as follows:
\begin{align}
\tilde \varepsilon_0 & = \varepsilon_0 + \varepsilon_1 \frac{M}{B_1}, \\
\tilde D & = \varepsilon_2 - \varepsilon_1 \frac{B_2}{B_1}, \\
\tilde A_2 & = A_2 \sqrt{1-\left( \frac{\varepsilon_1}{B_1} \right)^2}, \\
\tilde R_1 & = R_1 \sqrt{1-\left( \frac{\varepsilon_1}{B_1} \right)^2}.
\end{align}

The results for a fit in the region $k < 0.04~\angstrom^{-1}$ are shown in Table~\ref{tab:parameters} where we have chosen $\varepsilon_0 = -\varepsilon_1(M/B_1)$ which puts the Dirac point at zero energy. The bulk bands where then fitted to the bulk \emph{ab initio} calculation in a momentum sphere of radius $0.06~\angstrom^{-1}$ centered at the $\Gamma$ point under the constraints provided by the fit of the surface Dirac cone. The resulting bulk parameters are also shown in Table~\ref{tab:parameters}.
\begin{table}
	\centering
	\begin{ruledtabular}
	\begin{tabular}{lccc}
 	$a$ (\AA) & $c$ (\AA) & $u$ & $v$ \\
	\hline
	4.138 & 28.64 & 0.399 & 0.206 \\
	\end{tabular}
	\end{ruledtabular}
	\caption{\label{tab-a} Structural parameters of Bi$_2$Se$_3$ where $a$ and $c$ are the lattice parameters 
of the hexagonal cell, and $u$ and $v$ are the internal parameters of the rhombohedral axes \cite{Madelung1998}.}
	\label{tab:dft}
\end{table}

\section{Results and Discussion}
\label{sec:results}

\subsection{Results}

Here, we present our results for the interface spectrum of the TI junction shown in Fig.~\ref{fig:junction}. The spin chirality of the surface states, or equivalently, the mirror Chern number is determined by $\sign \left( A_1 A_2 \right)$ \cite{Liu2010}. The mirror Chern number is explicitly calculated in Appendix \ref{appendix:chern}. However, since the energy fitting is insensitive to the sign of $A_1$ and $A_2$, as can be seen from the bulk spectrum given in \eqref{eq:bulkbands}, we manually change the spin chirality of TI2. In this way we create a model for an interface between two $\BiSe$-like TIs with opposite mirror Chern number using realistic parameters. This can be done in two ways which correspond to different coupling regimes. We find that changing the sign of $A_2$ corresponds to strong coupling, while changing the sign of $A_1$ gives a weak coupling between the topological surface states \cite{DeBeule2013}. In Ref.~\onlinecite{Liu2010} it is shown that $A_1$ does not depend on the sign of the SOC constant so that we only consider the strong coupling regime. In the following, we always take $A_2^{(2)} = -A_2^{(1)}$, while the other parameters are equal and the parameters for TI1 are given in Table~\ref{tab:parameters}.

\begin{figure}
\centering
\includegraphics[width=.8\linewidth]{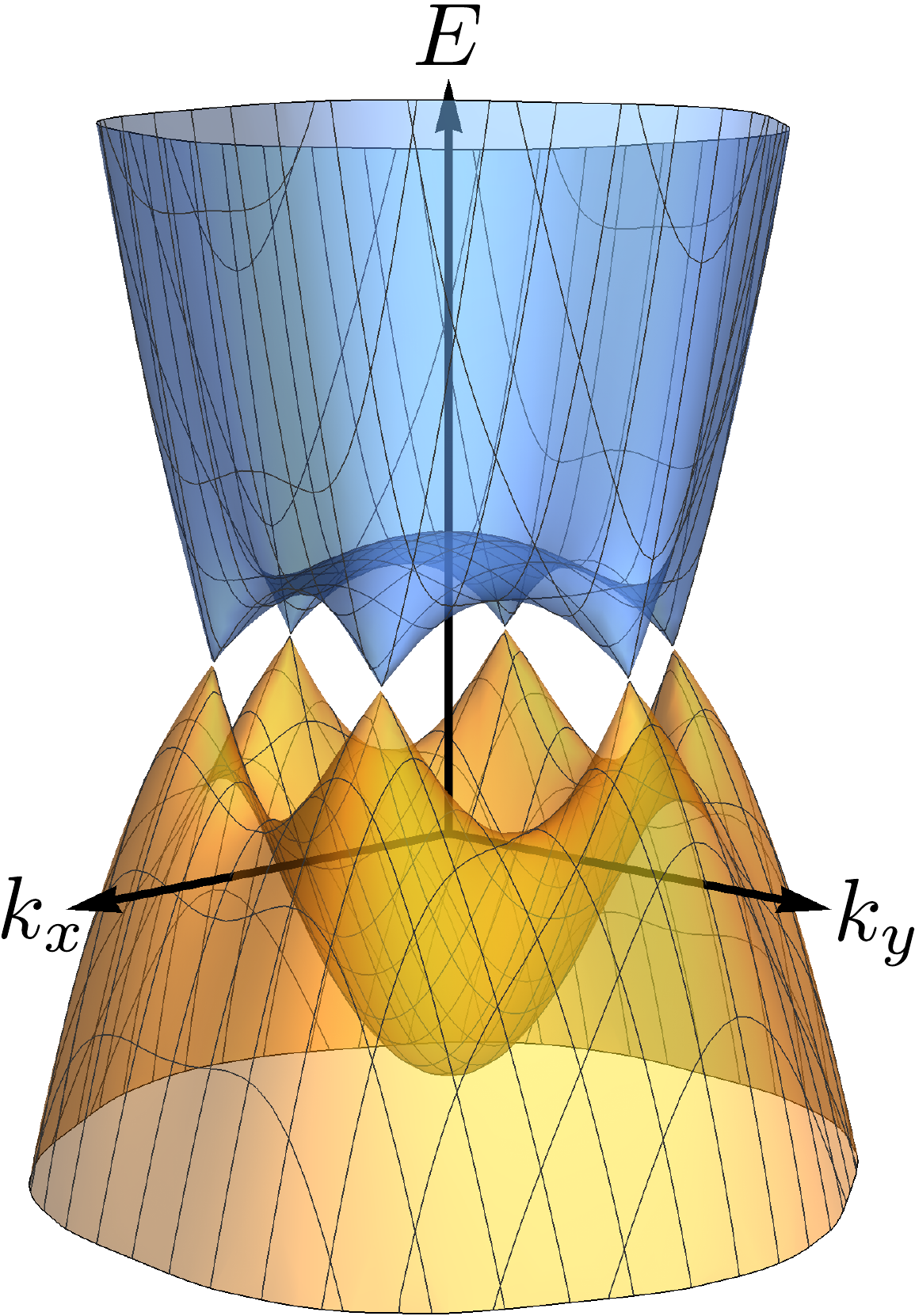}
\caption{Spectrum of interface states in the 2D interface Brillouin zone with $\bm k \in \left[ 0, 0.15 \right] \times \left[ 0, 0.15 \right]$ in units $\angstrom^{-1}$.}
\label{fig:spectrum}
\end{figure}

\begin{figure}
\centering
\includegraphics[width=\linewidth]{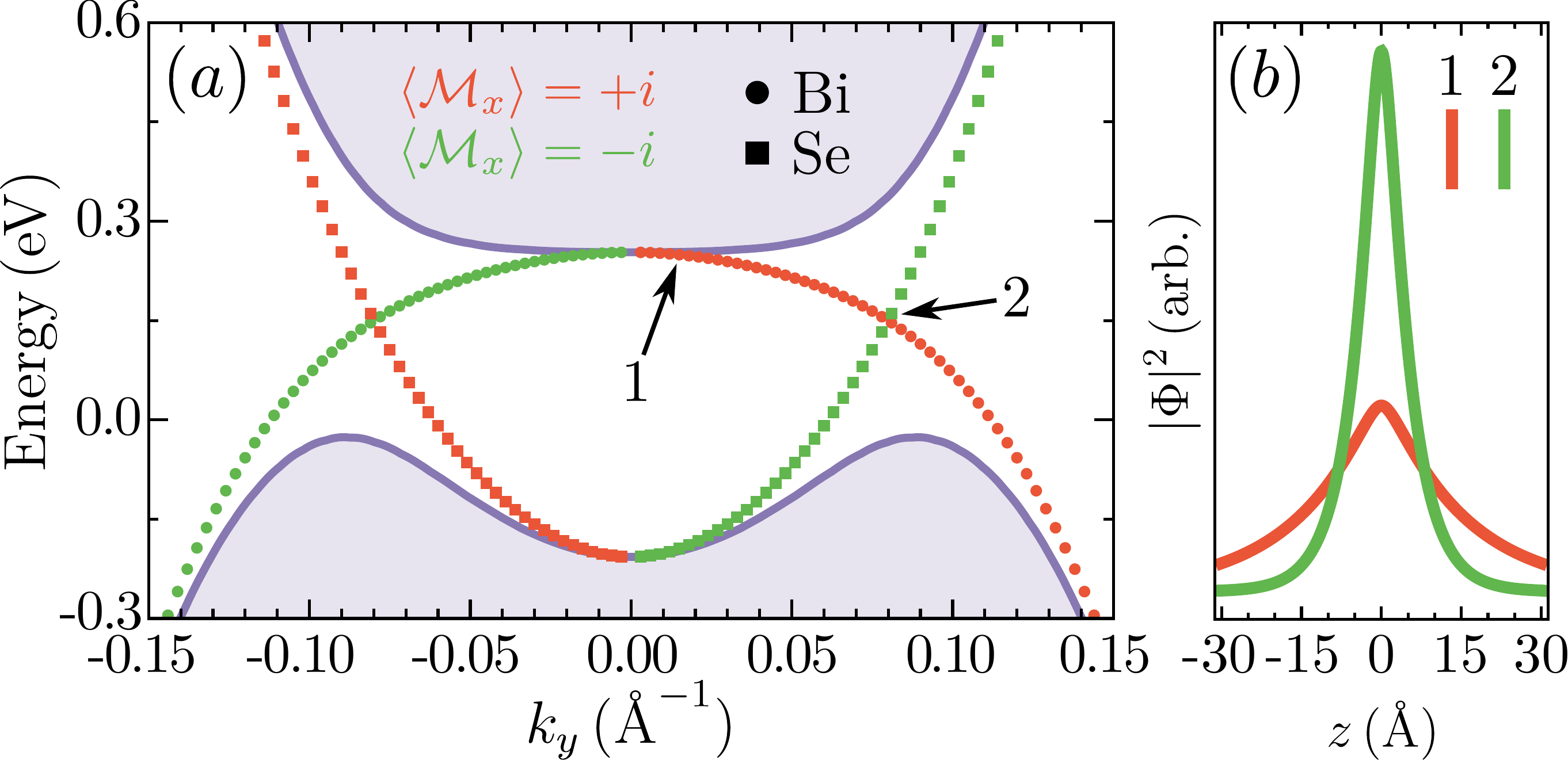}
\caption{(Color online) $(a)$ Spectrum of interface states along the mirror-symmetric line $k_x=0$ together with the projected bulk bands. Here, we show the mirror eigenvalues $\left< \Phi \right| \mathcal M_x \left| \Phi \right> = \pm i$ as red and green, respectively, and the orbital polarization as the dots (Bi) and squares (Se). $(b)$ Probability density of the interface states marked in $(a)$.}
\label{fig:mirror}
\end{figure}

In Fig.~\ref{fig:spectrum}, we show the interface spectrum which consists of six anisotropic Dirac cones with Dirac points located at $\left\{ (0, \pm k_0), (\pm \sqrt{3}k_0/2, \pm k_0/2) \right\}$ with $k_0 \approx 0.08~\angstrom^{-1}$. The six Dirac points lie on the three mirror axes of the 2D interface Brillouin zone, related by $C_3$ symmetry, and are protected by mirror symmetry. For example, the Dirac points $\left\{ (0, \pm k_0 \right\}$ lie on the mirror line $k_x=0$ and are protected by $\mathcal M_x$. Time reversal makes the Dirac cones symmetric with respect to $\bm k \rightarrow -\bm k$ but they would survive even if time reversal is broken as long as the mirror symmetry is preserved. Away from the mirror axes, the interface states are not protected and a gap is opened due to the cubic warping terms since lower order terms have full rotation symmetry. We now demonstrate that the interface states are protected by mirror symmetry. For $k_x=0$, $\mathcal M_x$ commutes with the Hamiltonian \eqref{eq:ham} so that at $k_x=0$, the interface states are eigenstates of $\mathcal M_x$. Hence, we show the spectrum along $k_x = 0$ together with the corresponding mirror eigenvalues in Fig.~\ref{fig:mirror}~$(a)$. We observe that the level crossings in the interface spectrum are protected because the corresponding branches of interface states have opposite mirror eigenvalues. Hence, the interface states remain gapless as long as the mirror symmetry is preserved. We also see that there are always two Kramers pairs at each energy in the gap so that the interface states, unlike the topological surface states, are not stable against disorder even if time reversal is preserved. The orbital polarization of the interface states is also shown and we find that the states are either completely localized on Bi (dots) or Se (squares). Note that there are actually four orbital characters: Bi and Se for TI1 ($\BiSe$) and two other ones for TI2. However, since we have chosen equal parameters for TI1 and TI2 (except for the spin chirality), the orbital structure of TI1 and TI2 coincide. Furthermore, we show the probability density of the interface states in Fig.~\ref{fig:mirror}~$(b)$. We see that the density of the interface states is spread over the entire junction. Moreover, at the crossing point, the density is localized more at the junction, and it spreads out more as the interface states merge with the bulk bands since the decay length diverges as the states approach the projected bulk bands. Note that the density is smooth at the interface ($z=0$) because, for our parameters, the continuity of the probability current density \eqref{eq:current} reduces to the continuity of the wave function and its derivative. 


\subsection{Scattering paradox}

\begin{figure}
\centering
\includegraphics[width=.85\linewidth]{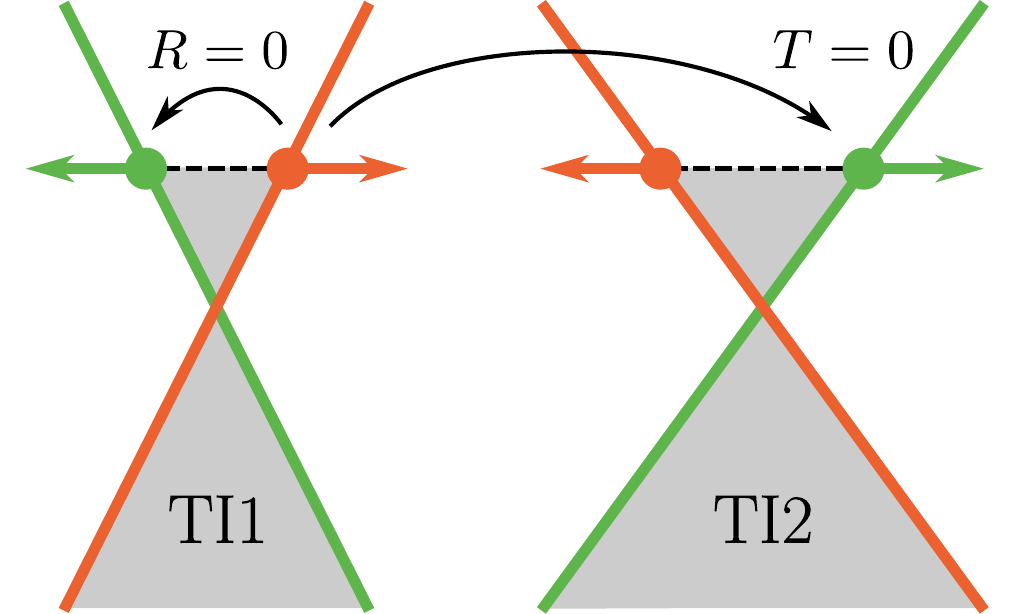}
\caption{(Color online) Scattering of surface states with opposite spin chirality on the $xz$ surface at the interface between TI1 and TI2 at normal incidence ($k_x=0$). In this case the mirror symmetry $\mathcal M_x$ about the $yz$ plane ensures that the spin is locked perpendicular to the momentum with $s_x = \pm$ (green and red). Both reflection ($R$) and transmission ($T$) are forbidden because $s_x$ is conserved. The dashed lines indicate the Fermi energy.}
\label{fig:scattering}
\end{figure}

The existence of the gapless interface states can also be understood by considering scattering of the topological surface state on the $xz$ surface at the interface between TI1 and TI2 \cite{Takahashi2011}. Mirror symmetry $\mathcal M_x$ enforces that the spin of the surface state on the $xz$ (or $xy$) surface is locked perpendicular to the momentum for $k_x=0$ which corresponds to a surface state that propagates in the $z$ (or $y$) direction. Indeed, we have $[H(k_x=0),\mathcal M_x]=0$ with $\mathcal M_x = -is_x$. Now consider a right-moving mode on the $xz$ surface of TI1 that scatters at the interface with TI2, as illustrated in Fig.~\ref{fig:junction}. At normal incidence ($z$ direction), we have $k_x = 0$, so that $s_x$ is conserved due to the mirror symmetry $\mathcal M_x$. However, in case the spin chirality of the surface states of TI1 and TI2 is opposite, neither reflection or transmission conserves $s_x$, which is illustrated in Fig.~\ref{fig:scattering}. This paradox is resolved if the incoming state can scatter into the interface, so that there must exist helical gapless states localized at the $xy$ interface for $k_x = 0$. These interface states arise from coupling of the topological surface states of TI1 and TI2. If the spin chirality is opposite, the overlapping surface bands at $k_x=0$ have opposite $s_x$, so that they are prevented from opening a gap. In general, gapless interface states exist only if the interface preserves the mirror symmetry, so they are not robust against disorder as is the case for the $\mathbb Z_2$ topological surface states.

\subsection{Rotational mismatch}

\begin{figure}
\centering
\includegraphics[width=.85\linewidth]{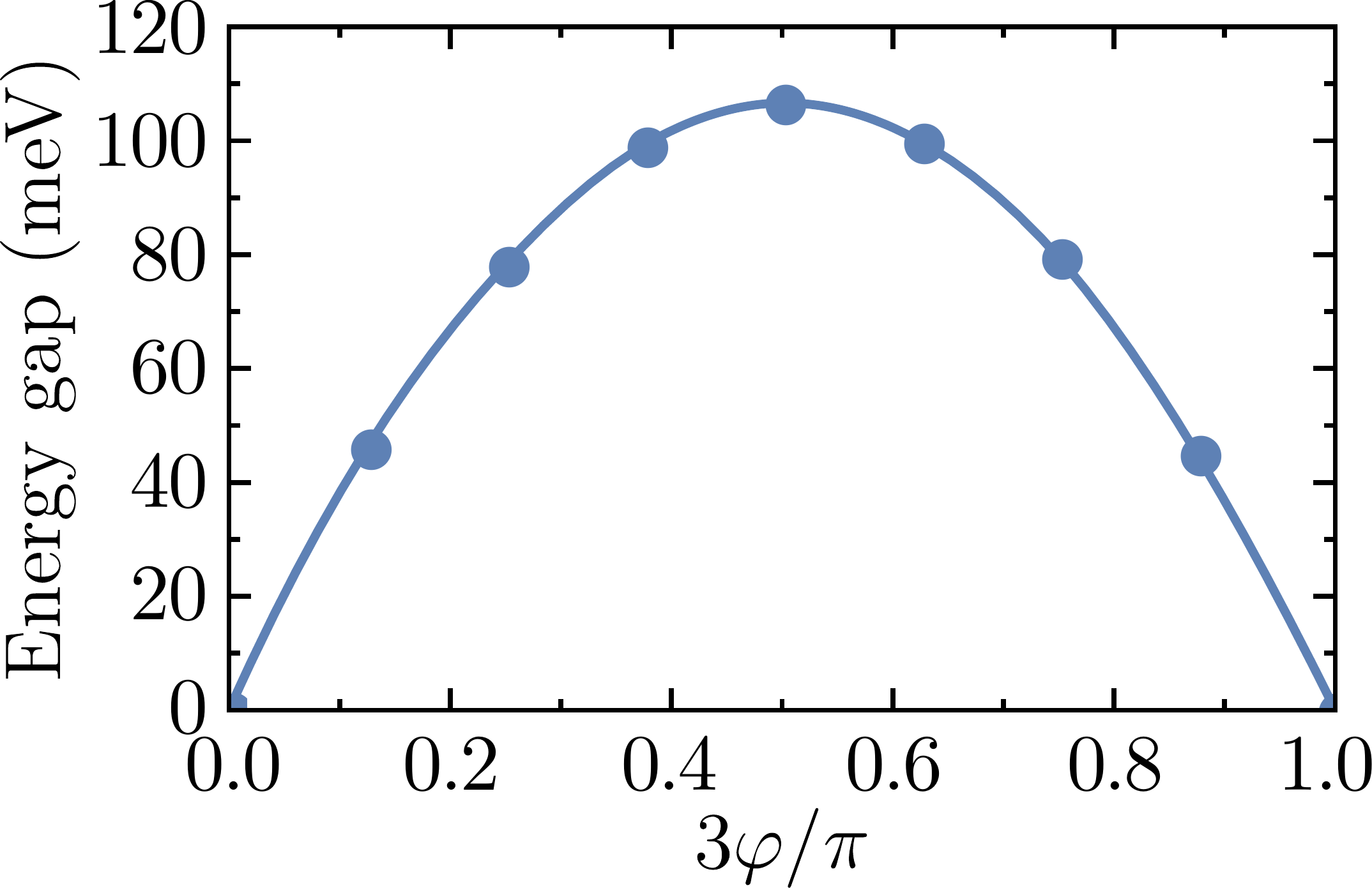}
\caption{Energy gap of the interface spectrum as a function of the rotational mismatch angle $\varphi$. The curve is interpolated between the calculated points (dots).}
\label{fig:mismatch}
\end{figure}

We also investigate the effect of rotational mismatch between the two TIs which breaks the mirror symmetry and thus opens a gap in the interface spectrum. The action of a rotation under an arbitrary angle $\varphi$ on the Hamiltonian is given by
\begin{equation}
H' = e^{-i\frac{\varphi}{2}\sigma_z} H(\bm k',k_z) e^{i\frac{\varphi}{2}\sigma_z} = H_0 + H_3',
\end{equation}
where $\bm k' = R(-\varphi) \bm k$. The rotation has no effect on $H_0$ since it preservers the full rotation symmetry. On the other hand, we have $H_3' = H_3(k,\theta_{\bm k}-\varphi)$ which can also be written as
\begin{equation}
H_3' = H_3(\bm k) \cos 3\varphi + V(\bm k) \sin 3\varphi,
\end{equation}
where
\begin{equation}
V = \frac{R_1}{2i} \left( k_+^3 - k_-^3 \right) \tau_y - \frac{R_2}{2} \left( k_+^3 + k_-^3 \right) \sigma_z \tau_x.
\end{equation}
which anticommutes with the mirror operator $\mathcal M_x = -i \sigma_x \tau_z$. The Hamiltonians of the two TIs with the rotational mismatch can then be written as
\begin{align}
H^{(1)} & = H_0^{(1)} + H_3^{(1)}(k,\theta_{\bm k}), \\
H^{(2)} & = H_0^{(2)} + H_3^{(2)}(k,\theta_{\bm k}-\varphi).
\end{align}
Hence, the interface spectrum in the presence of rotational mismatch over an angle $\varphi$ can be calculated in the same way as before with the substitution $\theta_{\bm k} \rightarrow \theta_{\bm k}-\varphi$ in all expressions relating to TI2. In this way, we numerically obtain the energy gap induced by rotational mismatch at the interface, which is shown in Fig.~\ref{fig:mismatch} as a function of the rotational mismatch angle $\varphi$. As expected, we find that the energy gap has period $\pi/3$ and that it attains a maximum of approximately $107$~meV at $\varphi = \pi/6$ when the rotational mismatch is maximal. It is clear that the magnitude of the energy gap depends on the parameters $R_1$ and $R_2$ and on the location of the crossing point $k_0$. The latter is understood since the cubic terms responsible for the gap are of the order of $k_0^3$ at the gap opening point.

\section{Summary}
\label{sec:summary}

In conclusion, we have investigated a junction of $\BiSe$-like TIs with opposite mirror Chern numbers. We used an effective continuum model, where the model parameters were obtained from \emph{ab initio} calculations, to calculate the spectrum of the interface states in a realistic model. We find that the gap closes at six Dirac points according to the mirror symmetry and $C_3$ symmetry. Moreover, we have explicitly shown that the interface states are protected by mirror symmetry by calculating the mirror eigenvalues along the mirror-symmetric line of the 2D interface Brillouin zone. The Dirac points always come in pairs and therefore the interface states are not protected by TR symmetry, unlike the topological surface state. Furthermore, we considered the effect of rotational mismatch, which breaks the mirror symmetry, on the interface states. We found that an energy gap is opened in the interface spectrum which has a  period $\pi/3$ as a function of the mismatch. 

A possible way to find signatures of these interface states is by applying a magnetic field along the $x$ direction (mirror axis), which does not break the mirror symmetry $\mathcal M_x$ but does break the $C_3$ symmetry in case of $\BiSe$-like TIs. This destroys four of the six cones and one could measure the conductance through the interface which should drop by a factor of $3$ when the magnetic field is applied. Moreover, if the sample is rotated around the $z$ direction, the conductance oscillates with a period of $\pi/3$ as a function of the rotation angle. This also gaps the Dirac cone on the transverse surface, which should enhance the signature.

Further studies are required to identify candidate materials for experimental realizations. To this end, one must identify commensurate strong TIs whose surface states have opposite spin chirality, and moreover, whose band gaps overlap in energy. Most likely, commensurability requires that these materials come from the same family of TIs, which is not necessarily the $\BiSe$ family. First-principles tools can then be used to model an interface using two slabs of these TIs.

\appendix

\section{Bulk spectrum}
\label{appendix:bands}

The bulk energy spectrum of \eqref{eq:ham} is doubly degenerate due to the combination of time-reversal symmetry and space inversion symmetry and given by
\begin{widetext}
\begin{equation}
\label{eq:bulkbands}
\begin{split}
& E_\pm(\bm k, k_z) = \varepsilon_0 + \varepsilon_1 k_z^2 + \varepsilon_2 k^2 \\
& \quad \pm \sqrt{\left( A_2 k \right)^2 + \left( M - B_1 k_z^2 - B_2 k^2 \right)^2 +k^6 \left( R_1^2 \cos^2 3\theta_{\bm k} + R_2^2 \sin^2 3\theta_{\bm k} \right) + A_1 k_z \left( A_1 k_z + 2 R_2 k^3 \sin 3\theta_{\bm k} \right)},
\end{split}
\end{equation}
\end{widetext}
with $\theta_{\bm k} = \arctan \left( k_y / k_x \right)$. 

\section{Expressions for $\lambda_\alpha$}
\label{appendix:lambda}

The expressions of the $\lambda_\alpha$ ($\alpha=1,2,3,4$) are given by the solutions of \eqref{eq:lambda}. To write down the explicit expressions, we first rewrite Eq.~\eqref{eq:lambda} as
\begin{equation}
a x^4 + b x^2 + c x + d = 0,
\end{equation}
with
\begin{align}
a & = D_1 D_2, \\
b & = A_1^2 + D_1 \left( E - L_2 \right) + D_2 \left( E - L_1 \right), \\
c & = i A_1 \left( N_1 + N_2 \right), \\
d & = \left( E - L_1 \right) \left( E - L_2 \right) - A_2^2 k^2 - N_1 N_2,
\end{align}
where the definitions of $D_{1,2}$, $L_{1,2}$, and $N_{1,2}$ are given in Section \ref{sec:localized}. The four roots $x_\alpha$ can then be written as
\begin{align}
x_{1,2} & = +S \pm \frac{1}{2} \sqrt{-4S^2 - \frac{1}{a} \left( 2b + \frac{c}{S} \right)}, \label{eq:xa} \\
x_{3,4} & = -S \pm \frac{1}{2} \sqrt{-4S^2 - \frac{1}{a} \left( 2b - \frac{c}{S} \right)}, \label{eq:xb}
\end{align}
where
\begin{align}
S & = \frac{1}{2} \sqrt{\frac{Q + \Delta_0/Q - 2b}{3a}}, \\
Q & = \sqrt[3]{\frac{\Delta_1 + \sqrt{\Delta_1^2-4\Delta_0^3}}{2}},
\end{align}
with
\begin{align}
\Delta_0 & = b^2 + 12 a d, \\
\Delta_1 & = 2 b^3 + 27 a c^2 - 72 a b d.
\end{align}
In case there are no imaginary solutions (in which case there would be no normalizable solutions), $\lambda_{1,2}$ are given by the two solutions from Eqs.~\eqref{eq:xa} and \eqref{eq:xb} with $\re x_\alpha > 0$ and $\lambda_{3,4} = -\lambda_{1,2}^*$.

\section{Mirror Chern number}
\label{appendix:chern}

The existence of the interface modes can also be understood from the mirror Chern number which is a weak topological invariant that gives an additional topological crystalline classification of topological insulators with mirror symmetry \cite{Teo2008}. Hence, the $\BiSe$ class of TIs are both strong TIs and topological crystalline insulators protected by mirror symmetry \cite{Rauch2014}. We now calculate the mirror Chern number and show that it corresponds to the spin chirality of the surface states.

For $k_x = 0$, the Hamiltonian \eqref{eq:ham} commutes with the mirror operator $\mathcal M_x = -i \sigma_x \tau_z$ and the energy bands are labeled with the mirror eigenvalues $\pm i$. The occupied mirror eigenstates are obtained by first finding an eigenstate $\left| \psi_1 \right>$ of one of the occupied bands at $k_x=0$. In this case, $\left| \psi_2 \right> = \mathcal M_x \left| \psi_1 \right>$ is also an eigenstate because $H(k_x=0)$ commutes with $\mathcal M_x$. The mirror eigenstates are then given by $\left| \phi_\pm \right> = \left| \psi_1 \right> \mp i \left| \psi_2 \right>$ since $\mathcal M_x \left| \phi_\pm \right> = \left| \psi_2 \right> \pm i \left| \psi_1 \right> = \pm i \left( \left| \psi_1 \right> \mp i \left| \psi_2 \right> \right)$ where we used $\mathcal M_x^2=-1$. In this way, we find that the normalized mirror eigenstates of the occupied bands are given by
\begin{widetext}
\begin{align}
\left| \phi_\pm(k_y,k_z) \right> & = \frac{1}{2 \sqrt{d \left( d + M - B_1 k_z^2 - B_2 k_y^2 \right)}}
\begin{bmatrix} A_1 k_z - R_2 k_y^3 \mp i A_2 k_y \\ -\left( d + M - B_1 k_z^2 - B_2 k_y^2 \right) \\ i A_2 k_y \mp \left( A_1 k_z - R_2 k_y^3 \right) \\ \mp \left( d + M - B_1 k_z^2 - B_2 k_y^2 \right) \end{bmatrix}, \\
d(k_y,k_z) & = \sqrt{\left( M - B_1 k_z^2 - B_2 k_y^2 \right)^2 + \left( A_1 k_z - R_2 k_y^3 \right)^2 + \left( A_2 k_y \right)^2}.
\end{align}
\end{widetext}
Since we consider $k_x=0$, the Hamiltonian is effectively two-dimensional and we can compute the Chern numbers of the mirror bands. The Chern number is defined as the integral over  the Berry curvature \cite{Thouless1982,Bernevig2013}. To obtain the Berry curvature we need the Berry connection
\begin{equation}
\bm A_\pm(k_y,k_z) = i \left< \phi_\pm \right| \nabla_{\left( k_y, k_z \right)} \left| \phi_\pm \right>,
\end{equation}
which can be written as $\bm A_\pm = \pm \bm A$ with
\begin{equation}
\begin{split}
\bm A & = \\
& \frac{\left( -A_2 \right) \left( M - B_1 k_z^2 - B_2 k_y^2 - d \right)}{2d \left[ \left( A_1 k_z - R_2 k_y^3 \right)^2 + \left( A_2 k_y \right)^2 \right]} \begin{bmatrix} A_1 k_z + 2 R_2 k_y^3 \\ -A_1 k_y \end{bmatrix}.
\end{split}
\end{equation}
The corresponding Berry curvature is then given by
\begin{align}
F_{yz} & = \partial_y A_z - \partial_z A_y \\
& = \frac{\left( -A_1 A_2 \right) \left( M + B_1 k_z^2 + B_2 k_y^2 + \frac{4B_1R_2}{A_1} k_y^3 k_z \right)}{2 d^ 3}.
\end{align}
We find that the mirror Chern numbers $n_\pm$ of the occupied bands $\left| \phi_\pm \right>$ are given by
\begin{align}
n_\pm & = \pm \frac{1}{2\pi} \int_{-\infty}^\infty \! dk_y \int_{-\infty}^\infty \! dk_z \, F_{yz} \\
& = 
\begin{cases}
\mp\sign \left( A_1 A_2 M \right) & \quad \mathrm{for}~M/B_{1,2} > 0 \\
0 & \quad \mathrm{for}~M/B_{1,2} < 0,
\end{cases}
\end{align}
where we verified the integral numerically. In accordance with time-reversal symmetry, the total Chern number of the occupied bands vanishes. However, the total mirror Chern number is nonzero in the inverted regime:
\begin{equation}
n_{\mathcal M} = \left( n_+ - n_- \right)/2 = -\sign \left( A_1 A_2 \right),
\end{equation}
for $M, B_1, B_2 > 0$ \cite{Teo2008}. The mirror Chern number $n_{\mathcal M}$ is a weak topological invariant protected by the mirror symmetry $\mathcal M_x$. Note that $n_{\mathcal M}$ is only defined on the three mirror planes in the Brillouin zone.

The gapless interface modes at $k_x=0$ can be understood from a change $\Delta n_{\mathcal M} = 2$ across the interface shown in Fig.~\ref{fig:junction}. The corresponding change in the Chern numbers $n_\pm$ gives rise to two left-moving and two right-moving modes in the $y$ direction \cite{Takahashi2011,Thouless1982,Bernevig2013}. This is similar to the surface states of the topological crystalline insulator SnTe, which has $n_{\mathcal M} = -2$ \cite{Hsieh2012}.
At a vacuum interface, the Chern numbers vanish and the change $\Delta n_\pm = \pm N$ leads to $N$ chiral and $N$ anti-chiral modes on the surface, respectively, or equivalently $N$ surface Dirac points. In general, the chiral and anti-chiral modes would annihilate pairwise, which is prohibited here by the mirror symmetry. Hence, the number of surface Dirac points is given by the absolute value $\left| n_{\mathcal M} \right|$ if the surface preserves the mirror symmetry. Moreover, in the presence of both time-reversal symmetry and mirror symmetry, the $\mathbb Z_2$ invariant is given by  $n_{\mathcal M}~\mathrm{mod}~2$ \cite{Teo2008}. 

\bibliographystyle{bibtex/mystyle.bst}
\bibliography{bibtex/paper.bib}

\end{document}